\documentclass{iopart}
\usepackage{graphicx}
\begin{document}

\title[Electrotransport and magnetic properies of Cr--GaSb spintronic materials]{Electrotransport and magnetic properies of Cr--GaSb spintronic materials synthesized under high pressure.}
\author{M V Kondrin$^1$, V R Gizatullin$^1$ , S V Popova$^1$,  A G Lyapin$^1$, V V Brazhkin$^1$, V Y Ivanov$^2$, A A Pronin$^2$, Y B Lebed$^3$ and R A Sadykov$^3$}
\address{$^1$ Institute for High Pressure Physics RAS, Troitsk, Moscow region, 142190 Russia}
\address{$^2$ General Physics Institute RAS, 117942 Moscow, Russia}
\address{$^3$ Institute for Nuclear Research RAS, Troitsk, Moscow Region, 142190 Russia}

\begin{abstract}
Electrotarnsport and magnetic properties of new phases in the system Cr--GaSb were studied. The samples were prepared by high-pressure (P=6--8 GPa) high-temperature treatment and identified by x-ray diffraction and scanning electron microscopy (SEM).  One of the CrGa$_2$Sb$_2$ phases with an orthorhombic structure $Iba2$ has a combination of ferromagnetic and semiconductor properties and is potentially promising for spintronic applications. Another high-temperature phase is paramagnetic and identified as tetragonal $I4/mcm$.
\end{abstract}
\submitto{\JPCM}
\section{Introduction}
In the recent years a lot of attention has ben given to the search for  spintronic materials capable of producing a spin-polarized electric current to carry signals modulated by small external magnetic fields. The problem of the generation of the spin-polarized electric current can be solved in two ways \cite{rmp:fert08} by either injecting an already polarized current into paramagnetic media or using the media where sustained polarized current could exist due to a difference between the density of the states for two spin orientations of carriers (in which case the material must be in the ferromagnetic state). Earlier the research effort in pursuing the latter goal was limited to searching for diluted magnetic semiconductors (i.e. semiconductors with a small -- about several percents -- amount of magnetic impurities), resulting in the synthesis of materials with the Curie temperature T$_C$ as high as 170 K in the case of (GaAs)$_{0.95}$Mn$_{0.05}$\cite{macdonald-2005-4,dietl-2002-17}. However, recently the research focus hass somewhat shifted to materials for which the theory predicts a full polarization of carriers, such as semiconductors with a zinc-blend structure \cite{jap:shirai03,prb:bose10}. For materials with a fully polarized current, the term ``half-metals'' was coined. CrSb and CrAs provide an example  of such a type of materials. Although these materials can be produced as ferromagnetic thin films \cite{li:jmmm07}, the study on their properties is hindered by the absence of bulk samples stable at normal conditions.

 The compound GaSb was previously widely used as a matrix for the construction of diluted magnetic semiconductors doped with manganese \cite{chen:511,gsm-beam-ep}. On the other hand chromium atoms (in comparison to Mn) form in A$^{3}$B$^{5}$ semiconductors energy levels close to the adge of valence or conduction bands. The manganese is extraodinary in this respect too as its energy levels lay deep into conduction band \cite{dietl-2002-17} so Cr is of little use to the production of diluted magnetic semiconductors, though first-principle calculations \cite{uspenskii:jmmm09} suggest a ferromagnetic nature of such solid solutions in the GaSb matrix.
    
Earlier, the attempts were made to use high-pressure synthesis for producing spintronic materials. In the studies \cite{en*gsm-synth,en*kondrin06:_gasb_mn,jopcs:kondrin08,jcsj:skakibara09} high pressure  was used to obtain a solution of manganese  in GaSb which has a zinc-blend structure under normal conditions. It should be noted that manganese is a popular dopant in the production of spintronic materials because, unlike  other magnetic atoms (Cr for example) it can much more easily form solid solutions in semiconductor lattices which leads to the formation of diluted magnetic semiconductors. In this article we present the data on the synthesis of magnetic and paramagnetic materials in the system Cr--GaSb under high pressure from 6 to 8 GPa.

\section{Synthesis and structure.}
The samples were obtained by quenching from the synthesis temperatures in the range 700-900 K to room temperature in Toroid anvils \cite{hpr:khvostantsev2004} under the synthesis pressures in the range P=6--8 GPa. The initial mixture consists of zinc-blend GaSb and a varying content of Cr in atomic proportion 1:$x$ where $x$=0.4--1. 

\begin{figure}
\includegraphics[width=\textwidth]{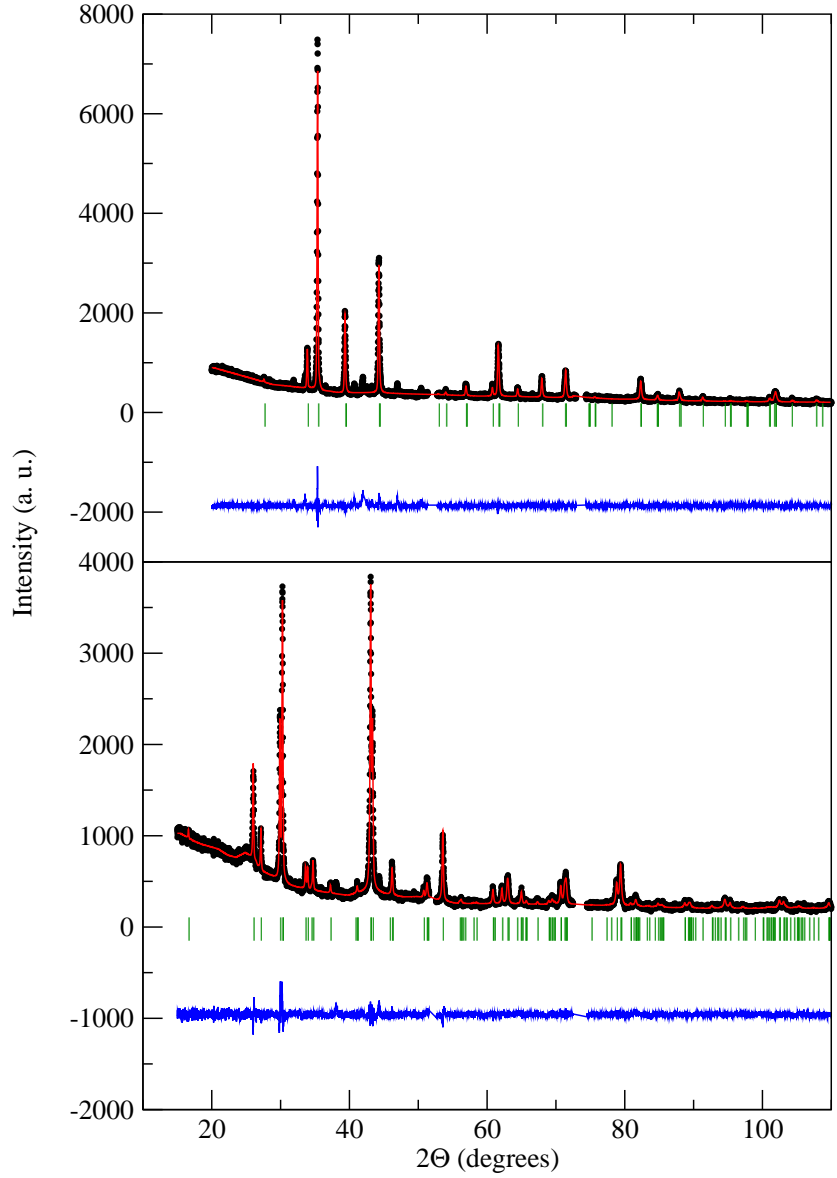}
\caption{X-ray diffraction data for two high-pressure phases: (a) CrGaSb (b)CrGa$_2$Sb$_2$}
\label{fig:x-ray}
\end{figure}

The X-ray diffraction measurements of the obtained phases were carried out with a Stoe STADI MP diffractometer device with a Cu$K_{\alpha}^{1}$ radiation source and curved germanium monochromator. It was found that, depending on the synthesis condition, two structurally different phases CrGa$_2$Sb$_2$ and CrGaSb were recovered with orthorhombic $Iba2$(No.45)\cite{Hahn2002} and tetragonal $I4/mcm$ (No.140)\cite{Hahn2002} space groups, correspondingly. The synthesis output was generally influenced by the proportion of the initial components and the synthesis temperature so the higher synthesis temperatures favors the formation of the latter phase.  The diffraction patterns of the two phases are displayed in Fig.~\ref{fig:x-ray}. According to X-ray diffraction phase analysis we have obtained two nearly pure phases: orthogonal for the Cr to GaSb proportions 0.5:1 and tetragonal for the propoertion 1:1, i.e. the initial composition corresponds to stoichiometric one\footnote{When our work was in its final stage we became aware of synthesis of the high-pressure CrGa$_2$Sb$_2$ phase by Sakakibara et al. \cite{sakakibara:jac10}}.

\begin{figure}
\includegraphics[width=\textwidth]{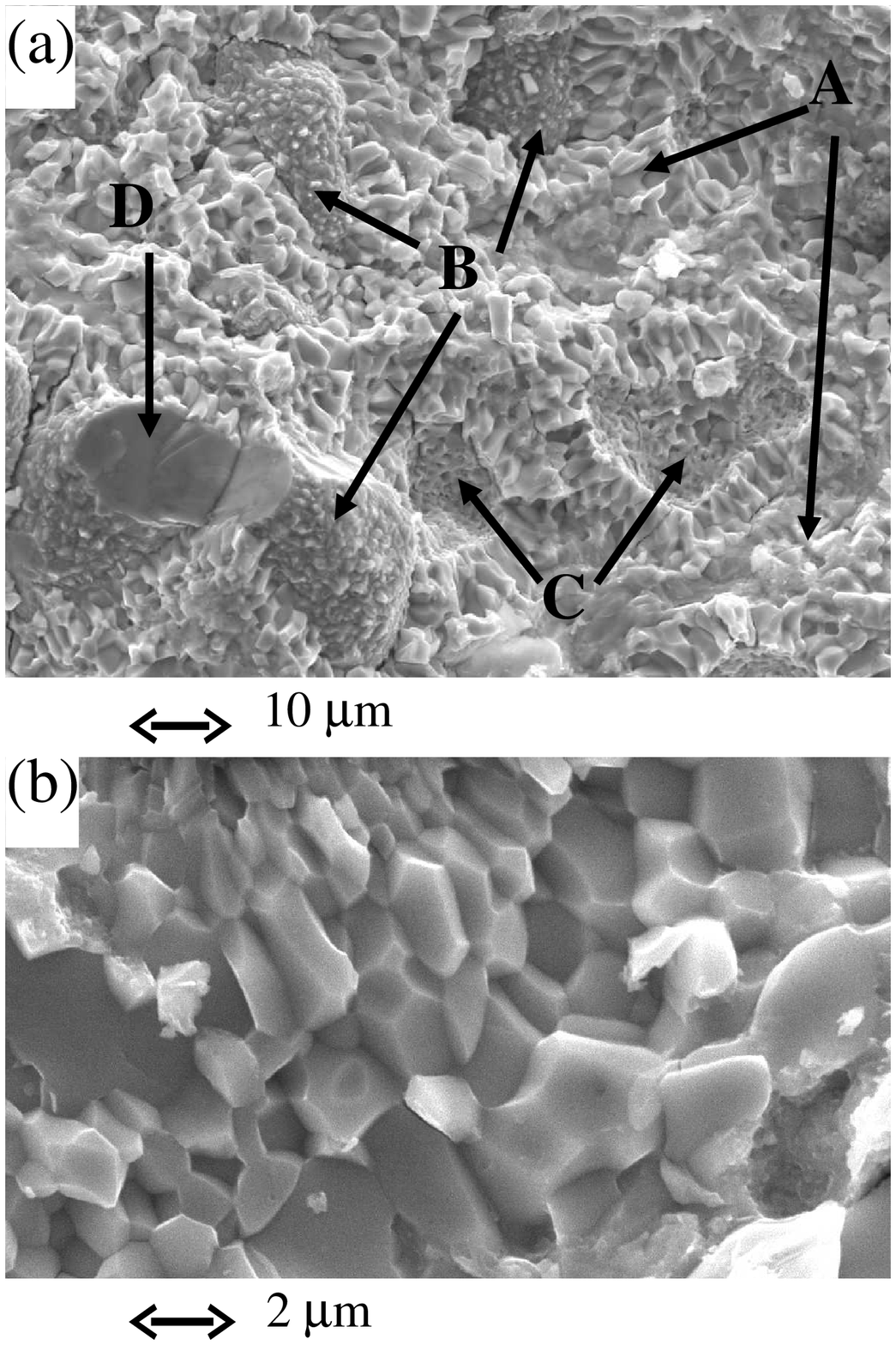}
\caption{SEM images of the high-pressure orthorhombic (a) and tetragonal (b) phases on the cleave of the samples. The following areas can be recognized from the elemental analysis in the panel (a): A is the prevailing area of the orthorhombic CrGa$_2$Sb$_2$ phase, B - grains of Cr, C - prints left by removed Cr grains on the CrGa$_2$Sb$_2$ phase, and D - shear of the Cr grain.}
\label{jeol}
\end{figure}

The morphology and chemical composition of the obtained samples were checked after the synthesis using a JEOL electron microscope (see images in Fig.~\ref{jeol}). For orthorhombic phase we specifically select the area with several Cr grains or their prints (Fig.~\ref{jeol}~(a)). The different areas in Fig.~\ref{jeol}~(a) were identified by elemental microanalysis using an energy dispersive spectrometer at the JEOL microscope. Th composition of the orthorhombic CrGa$_2$Sb$_2$ phase was found to be stoichiometric with the accuracy of 2-4 atomic \%. The observed morphology and composition picture for the orthorhombic phase indicates a diffusion mechanism of the formation of this phase. The SEM images from the tetragonal CrGaSb phase show a quite homogeneous morphology with typical picture
presented in Fig.~\ref{jeol}~(b). The elemental microanalysis gave a composition close to stoichiometric with several atomic \% excess of the Sb content.

It should be stressed that the samples consist practically uniformly of the single phase with point-like inclusions of surplus amount of chromium, but the chemical composition of the main phase is close to stoichiometric one. Chromium differs in this respect from analogous compounds with Mn atoms which according to Ref.~\cite{jopcs:kondrin08} form a series of solid-solutions in the orthorhombic phase (previously this phase in Refs.~\cite{en*gsm-synth,en*kondrin06:_gasb_mn,jopcs:kondrin08} was erroneously identified as a simple cubic structure).

For both phases the  full-profile Rietveld refinement was carried out using the Fullprof software \cite{pbcm:fullprof93}. The results are summarized in Tables \ref{tab:iba2} and \ref{tab:i4mcm}. The respective fits are also shown in Fig.~\ref{fig:x-ray}. The agreement factors for these fits are quite good: Bragg's R-factor is 4.43 for the orthorhombic phase and 6.21 for the tetragonal phase, which corroborates structural models suggested in \cite{en*gsm-synth,jcsj:skakibara09,sakakibara:jac10}. 

By using the lattice parameters we estimate the density of the new phases as 6.99 g/cm$^3$ ($Iba2$) and 7.3 g/cm$^3$ ($I4/mcm$), which is in good agreement with pycnometric densities measured using the same samples (7.0$\pm$0.02 and 7.3$\pm$0.02 g/cm$^3$ respectively). 

Although any extensive testing of the stability range of the obtained phases has not been provided we can conclude from the exothermal step-by-step annealing that both phases are metastable at normal conditions and decompose at room pressure at temperatures above 500 K.

\begin{table}
\caption{Crystal structure of the orthorhombic CrGa$_2$Sb$_2$ phase, space group  $Iba2$(No.45), $a$=11.772$\pm$0.007~\AA, $b$=5.964$\pm$0.007~\AA, $c$=5.897$\pm$0.007~\AA. Atomic positions, coordinates and site occupancies. }
\label{tab:iba2}
\begin{indented}
\item[]\begin{tabular}{@{}llllll}
\br
&Cite&X&Y&Z&Occ.\\
\mr
Ga&8(c)&0.130&0.269&0.243&1.00\\
Sb&8(c)&0.137&0.272&0.726&1.00\\ 
Cr&4(a)&0.000& 0.000&0.000&1.00\\
\br
\end{tabular}
\end{indented}
\end{table}

\begin{table}
\caption{Crystal structure of the tetragonal CrGaSb phase, space group  $I4/mcm$(No.140), $a$=6.466$\pm$0.007~\AA, $c$=5.291$\pm$0.007~\AA, $c/a$=0.82. Atomic positions, coordinates and site occupancies.}
\label{tab:i4mcm}
\begin{indented}
\item[]\begin{tabular}{@{}ccllll}
\br
&Cite&X&Y&Z&Occ.\\
\mr
Ga&8(h)&0.157&0.657& 0.000&0.589\\
Sb&8(h)&0.157&0.657&0.000&0.411\\ 
Cr&4(a)&0.000&0.000&0.250&1.000\\
\br
\end{tabular}
\end{indented}
\end{table}

\section{Magnetic Properties.}
\begin{figure}
\includegraphics[width=\columnwidth]{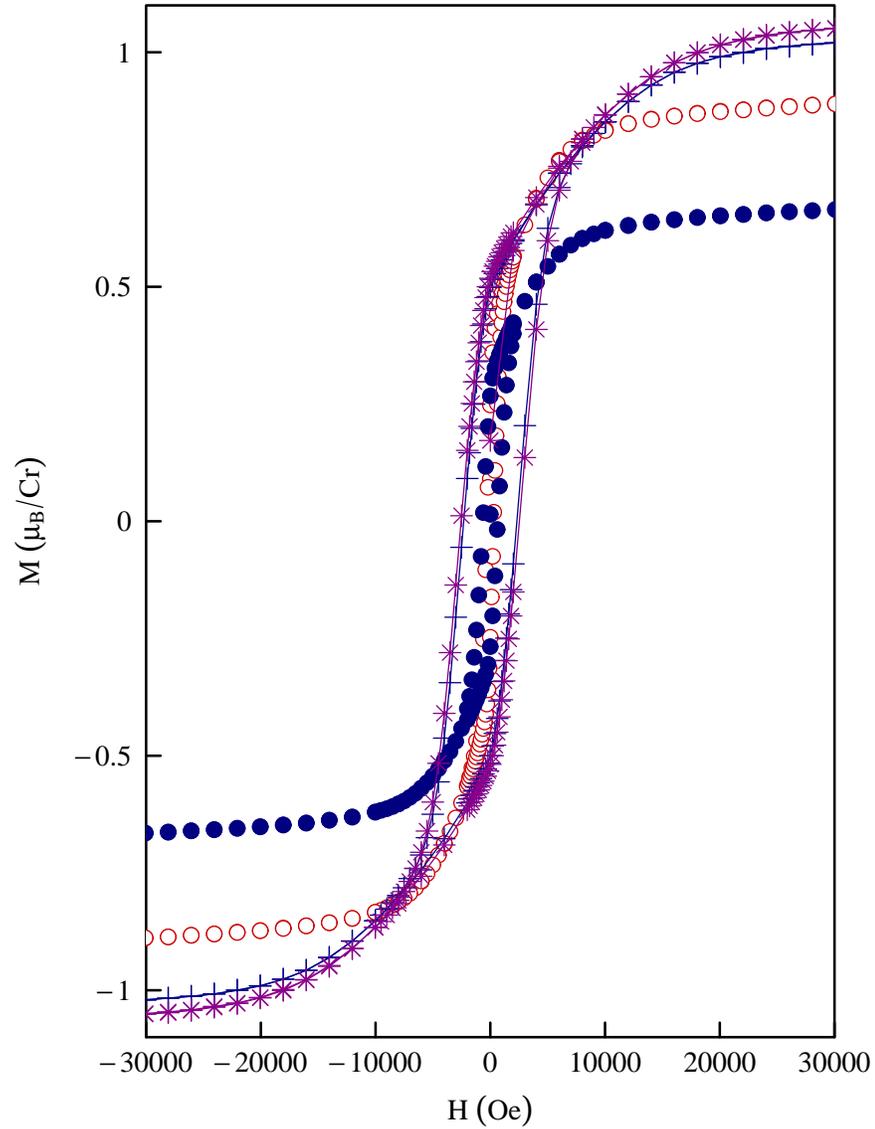}
\caption{Magnetization curves of CrGa$_2$Sb$_2$ samples with $Iba2$ structure with different chromium content (open symbols -- $x$=0.5, closed -- $x$=0.6) at different temperatures. $\circ$ -- T=290~K, $\bullet$ -- T=290~K, $+$ -- T=77~K, $\ast$ -- T=4.5~K}
\label{fig:mag_h}
\end{figure}

\begin{figure}
\includegraphics[width=\columnwidth]{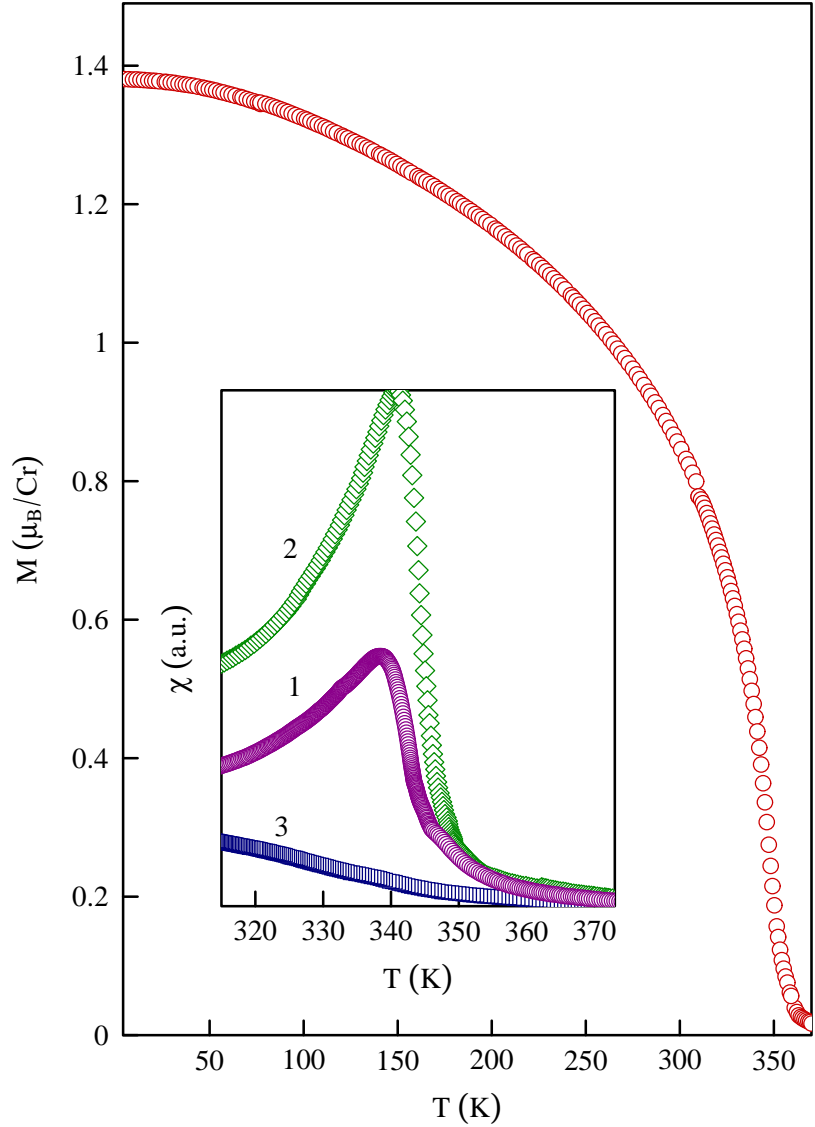}
\caption{Temperature dependence of the spontaneous magnetic moment in  CrGa$_2$Sb$_2$ (main panel) and magnetic susceptibility in the vicinity of the Curie temperature for the samples with the different chromium content (1 -- $x$=0.5, 2 -- $x$=0.6, 3 -- $x$=0.8) and same crystal structure $Iba2$(inset).}
\label{fig:mag_t}
\end{figure}

The magnetic moment was measured with a QuantumDesign SQUID magnetometer (fields up to 50~kOe); and sometimes we also use the data obtained with a  custom-made inductive device for the AC magnetic susceptibility measurements. The primary result of this research is the discovery of a ferromagnetic ordering in  CrGa$_2$Sb$_2$ samples with $Iba2$ structure. The magnetization curves are shown in Fig.~\ref{fig:mag_h}. It was found that the critical temperature T$_C$=350 K remains almost constant and slightly depends on the initial chromium content. Note that the critical temperature is substantially higher than the one for  MnGa$_2$Sb$_2$ which varies in the range 210--310 K depending on the Mn content \cite{en*gsm-synth,en*kondrin06:_gasb_mn,jopcs:kondrin08,jcsj:skakibara09}. On the other hand, an  increase in the chromium content up to x=0.8 leads to some smoothing and widening of the ferromagnetic transition as demonstrated in  Fig.~\ref{fig:mag_t} a) (compare the curve 3 with the curves 1-2). As seen from Fig.~\ref{fig:mag_h}, the spontaneous magnetic moment per Cr atom in the samples with higher chromium content (x=0.6) is below 1~$\mu_B$ at room temperature, which is slightly lower than in stochiometric phase. This discrepancy can be attributed to the fact that the excess amount of chromium may form antiferromagnetic precipitates in a ferromagnetic bulk sample, resulting in an apparent decrease in the magnetic moment. With a decrease in temperature, the spontaneous magnetic moment increases up to 1.5~$\mu_B$ per Cr atom at helium temperature (see Fig.~\ref{fig:mag_t}) which is accompanied by an increase in the magnetic coercive force (Fig.~\ref{fig:mag_h}).

\begin{figure}
\includegraphics[width=\columnwidth]{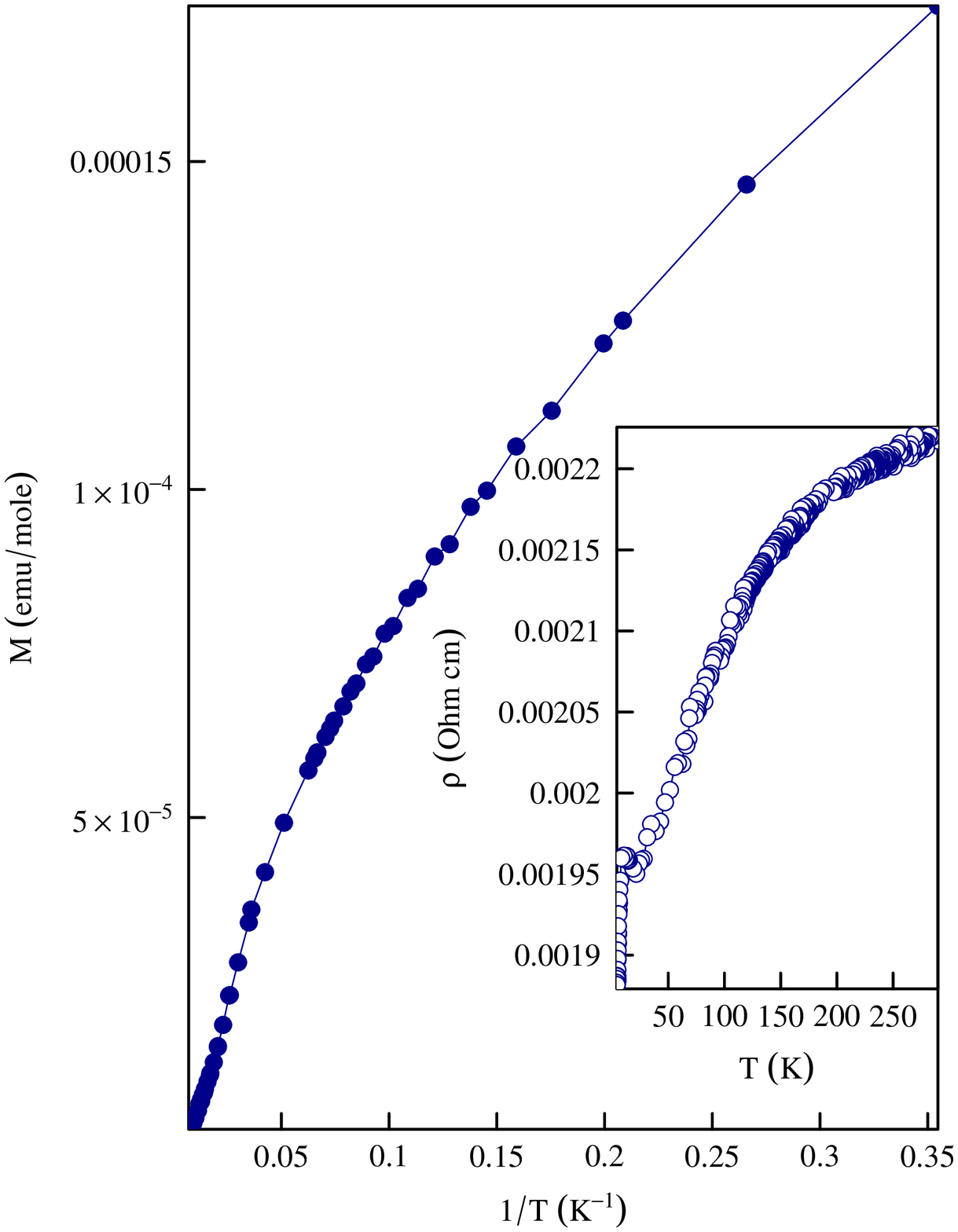}
\caption{Temperature dependence of the magnetic moment and resistivity (inset) of the tetragonal GaSbCr phase. }
\label{fig:paramagnetic}
\end{figure}

Unlike an analogous compound with manganese, the $I4/mcm$ phase CrGaSb has no magnetic ordering and is paramagnetic in temperature range T=4.5--300 K, as can easily be concluded   from the data in Fig.~\ref{fig:paramagnetic}.

\section{Electrotransport Properties.}
\begin{figure}
\includegraphics[width=\columnwidth]{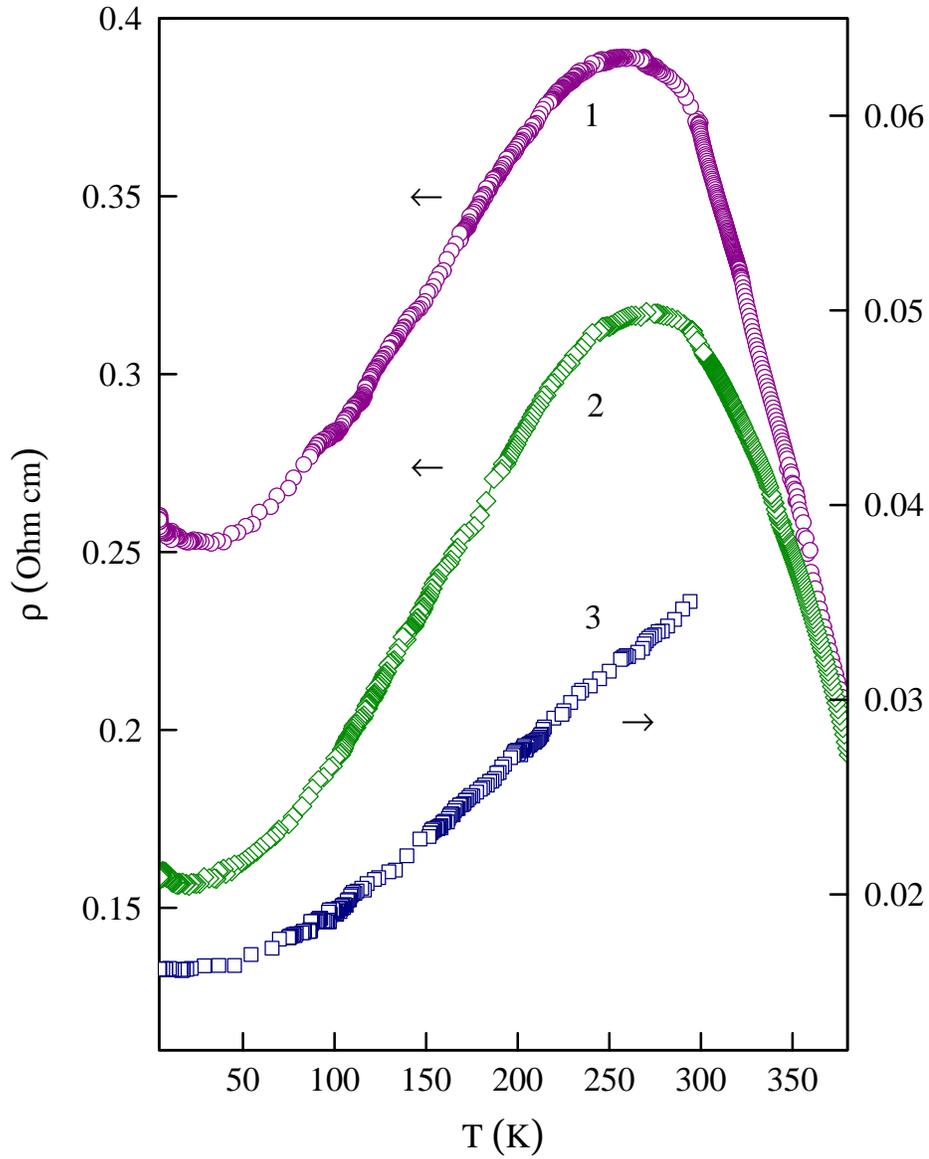}
\caption{Temperature dependence of the resistivity for the samples with the same crystal structure $Iba2$ and different initial chromium content (1 -- $x$=0.5, 2 -- $x$=0.6, 3 -- $x$=0.8)}
\label{fig:res}
\end{figure}

 The electrical resistance of the new Cr-GaSb phases, measured by 4-probe method, is much higher than that of the analogous manganese-based phases. Even the resistance of the metallic $I4/mcm$ phase (see inset in Fig.~\ref{fig:paramagnetic}) is almost order of magnitude higher than the values reported in Ref.~\cite{en*kondrin06:_gasb_mn} for the tetragonal phase MnGaSb. The elctrical resistance of the orthorhombic phase is a more complicated point because the bulk resistance of a pure phase can well be shunted by inclusions of metallic chromium. This may explain the observed decrease in the conductivity of $Iba2$ phases CrGa$_2$Sb$_2$ with the chromium content deviating from stochiometric (see Fig.~\ref{fig:res}). One should emphasize that the number of the point-like inclusions of Cr for near-stoichiometrical smaples (x $\sim$ 0.5) was very small. So we may consider the resistivity of the stochiometric phase ( Fig.~\ref{fig:res} curve 1) to be  intrinsic, while the deviation observed in curves 2-3 (Fig.~\ref{fig:res}) may by attributed to impurities. A small decrease of resistivity in the temperature range T=4.2--8.0 K observed for the tetragonal phase (see inset in Fig.~\ref{fig:paramagnetic}) can be attributed to gallium which, under certain conditions, can display superconducting properties \cite{jetpl:demishev92}.  

 The temperature dependence of resistivity of the orthorhombic phase can be devided in two temperature regions: the first one is  above the magnetic transition where the temperature dependence is similar to semiconducting one (i.e. increasing with decreasing temperature), and the second region below T$_C$ where the  resistance is almost independent of temperature. The transport properties of orthorhombic phase below below   T$_C$  can be compared with transport properties of ``dirty'' metals with strong impurity scattering.  However, in the case of CrGa$_2$Sb$_2$ (unlike MnGa$_2$Sb$_2$, as observed in Ref.~ \cite{jcsj:skakibara09}) there is no straightforward correspondence between the Curie temperature and the cusp in the resistivity curve because the temperature dependence of resistivity changes the trend at much lower temperatures than the temperature where the ferromagnetic ordering takes place (290~K and 350~K, respectively).  

\section{Conclusions.}

Better insight into the electronic state of both phases can be gained by comparing the observed resistivity with a calculated ``minimal metallic'' conductivity expected for these phases. Taking into account the lattice parameters, the upper limit for the conductivity in the mettalic state can be obtained with the assumption that a unit cell (of volume $V$) contribute to the conduction band one carrier with a scattering length equal to the shortest of lattice parameters ($\lambda$):
\begin{equation}
\sigma_{min}=(3\pi^2)^{-2/3}(e^2/\hbar)\frac{\lambda}{V^{2/3}}
\end{equation}
The calculation according to the above formula yields the values $\rho_{max}$=$1/\sigma_{min}$=3.7$\cdot$10$^{-3}$ Ohm$\cdot$cm and $\rho_{max}$=2.7$\cdot$10$^{-3}$ Ohm$\cdot$cm for the orthorhombic and tertragonal phases respectively. This means that tetragonal phase is almost on the verge of a metal-insulator transition while the orthorhombic phase is in the semiconducting  state with a typical resistivity almost two order of magnitude higher than the evaluated maximum metallic resistivity. Note that the presence of conductive inclusions in the samples doesn't contradict the above conclusion because the impurities do increase the measured conductivity of the samples but this conductivity is still well below the $\sigma_{min}$.

Although the performed measurements we have carried out couldn't provide information about polarization of carriers in the samples, we believe that the  CrGa$_2$Sb$_2$ compound may prove to be an interesting candidate for spintronic applications because of its fairly high Curie temperature and semiconducting properties.
\ack 
The work has been supported by the grants of RFBR (08-02-00014 and 10-02-01407) and by the Programs of the Presidium of RAS. The authors are grateful for O.A.Sazanova for discussion and useful suggestions.
\section*{References}
\bibliographystyle{unsrt}
\bibliography{crgasb}
\end{document}